\documentclass{aa}
\usepackage{psfig}
\usepackage{graphicx}

\def\secondip{\hbox{\rlap{\hbox{.}}\hbox{$''$}}}

\def\gradip{\hbox{\rlap{\hbox{.}}\raise 5.truept \hbox{{\small $\circ$}}}}

\begin{document}

\title{HST NICMOS Photometry of the reddened bulge globular clusters 
NGC 6528, Terzan 5, Liller 1, UKS 1 and Terzan 4
\thanks{
Based on observations collected with the NASA/ESA {\it Hubble Space
Telescope} obtained at the Space Telescope Science Institute, which is
operated by the Association of Universities for Research in Astronomy,
Inc., under NASA contract NAS 5-16555}}

\author{S. Ortolani \inst{1}, B. Barbuy\inst{2} E. Bica\inst{3}, A.
Renzini\inst{4}, M. Zoccali\inst{4}, R.M. Rich\inst{5}, S. Cassisi\inst{6} }

\offprints{S. Ortolani}

\institute{
Universit\`a di Padova, Dept. di Astronomia, Vicolo dell'Osservatorio 2, 
I-35122 Padova, Italy; ortolani@pd.astro.it
\and
Universidade de S\~ao Paulo, CP 3386, S\~ao Paulo 01060-970, Brazil;
barbuy@iagusp.usp.br
\and
Universidade Federal do Rio Grande do Sul, Dept. de Astronomia, CP 15051, 
Porto Alegre 91501-970, Brazil;
bica@if.ufrgs.br
\and
European Southern Observatory, Karl Schwarzschild Strasse 2, D-85748, 
Garching bei M\"unchen, Germany; arenzini@eso.org, mzoccali@eso.org
\and
Department of Physics and Astronomy, Division of Astronomy and Astrophysics, 
University of California, Los Angeles, CA 90095-1562; rmr@astro.ucla.edu
\and
Osservatorio Astronomico Collurania, I-64100, Teramo, Italy; 
cassisi@astrte.te.astro.it
}

\date{}
\authorrunning{S. Ortolani et al.}
\titlerunning{HST NICMOS CMDs of bulge clusters}

\abstract{
We present results from NICMOS Hubble Space Telescope observations of
the reddened bulge globular clusters NGC~6528, Terzan~5, Liller~1,
UKS~1 and Terzan~4, obtained through the filters F110W (1.1 $\mu$m)
and F160W (1.6 $\mu$m) (nearly equivalent to $J$ and $H$). For the
first time the turnoff region of Liller~1 and the main sequence of
Terzan~5 and Terzan~4 are reached, as well as the horizontal branch of
UKS 1.  The magnitude difference between the turnoff and the red
horizontal branch $\Delta m_{110}=m_{110}^{\rm HB}- m_{110}^{\rm TO}$
is used as an age indicator. From comparisons with new isochrones in
the NICMOS photometric system, we conclude that the two metal-rich
clusters NGC~6528 and Terzan~5 are coeval within uncertainties ($\sim
20\%$) with 47~Tucanae.  Liller~1 and UKS~1 are confirmed as
metal-rich globular clusters. Terzan~4 is confirmed as an interesting
case of a metal-poor cluster in the bulge with a blue horizontal
branch.
}

\maketitle

\section{Introduction} 

\begin{table*}
\begin{center}
\caption[1]{Reddening and metallicity of the program clusters}
\begin{tabular}{lrrcrcr}
\hline
\noalign{\smallskip}
 &  &  & \multispan2 Barbuy et al. 1998 & \multispan2 Harris' compilation \\
\noalign{\smallskip}
Cluster & $l$ & $b$ & $E(B-V)$ & [M/H] & $E(B-V)$ & [M/H]   \\
\noalign{\smallskip}
\hline
\noalign{\smallskip}
Terzan 5  & $3.81$   & $1.67$  & 2.39  & $0.00$  & 2.37 & $-0.28$  \\
NGC~6528  & $1.14$   & $-4.18$ & 0.52  & $-0.2$  & 0.56 & $-0.17$  \\
Terzan 4  & $356.02$ & $1.31$  & 2.31  & $-2.0:$ & 2.35 & $-1.60$  \\
Liller 1  & $354.81$ & $-0.16$ & 3.00  & $+0.2$  & 3.00 & $+0.22$  \\
UKS 1     & $5.13$   & $0.76$  & 3.10  & $-0.5$  & 3.09 & $-0.50$  \\
\noalign{\smallskip} \hline \end{tabular}
\end{center} 
\end{table*}

Because they are simple stellar populations of single age and
abundance, globular clusters located in the central regions of the
Galaxy offer a special opportunity to study the stellar populations in
the Galactic bulge (e.g. Minniti 1995; Ortolani et al. 1995; Barbuy et
al. 1998).  The distribution of cluster metallicities coupled with
their spatial location places interesting constraints on the formation
processes and timescales of the inner spheroid (Ortolani et al. 1995
and references therein).  The kinematics of globular clusters within 4
kpc from the Galactic center (as derived from the radial velocities of
cluster members) indicates that most of such clusters do indeed belong to
the bulge (C\^ot\'e 1999). Fundamentally, we want to know the ages of these
clusters relative to their counterparts in the distant halo, and
relative to the field stars in the Galactic bulge.

Many of the bulge globular clusters suffer from significant extinction
and some, such as Liller~1, are essentially invisible in the optical.
However, even with ground-based infrared detectors, study of these
clusters has been difficult: they are extremely crowded, and located
in the heavily crowded bulge field.  Their great distances have
prevented efforts to reach the turnoff point from the ground.  The
extinction is large and spatially variable, so that even in the
infrared the main sequences of their color-magnitude diagrams
(CMDs) have large scatter.  For those clusters in the inner bulge,
very high spatial resolution and point spread function stability is
required for success, hence our choice to use the NICMOS detector on
board HST.

Ortolani et al. (1995) have analysed the relatively low reddening
bulge clusters NGC~6528 and NGC~6553 using HST optical $V$ and $I$
photometry, and concluded that their age is close to that of the inner
halo cluster 47 Tucanae. NGC 6553 has also been studied in the near-IR
($J$ and $K$ photometry, Guarnieri et al. 1998), while an HST proper
motion study of stars in its field has allowed to largely
decontaminate its CMD from foreground/background stars belonging
either to the disk or the bulge (Zoccali et al. 2001a); the resulting very 
clean turnoff point confirms the cluster age. 

Terzan~1, a highly reddened globular cluster well within the bulge,
has also been studied with HST (Ortolani et al. 1999), revealing that
this is a low metallicity cluster with a fairly red
horizontal branch, i.e. a {\it second parameter} cluster, possibly a
few Gyr younger than the majority of globular clusters in the Galaxy.

In the present study we further expand these studies, providing new
results based on HST/NICMOS observations of NGC 6528 and of four other
among the most reddened bulge and lowest galactic latitute clusters,
namely UKS 1, Liller 1, Terzan 4 and Terzan 5. Near-IR observations
are the best hope for deriving deep CMDs for these clusters.  Even so, 
we are able to report ages for only 2 of the four clusters.

Reddening and metallicities for the sample clusters (Table~1) have
been taken from Barbuy et al. (1998), and the compilation by Harris
(1996), as updated in the web page
http://phy\-sun.phy\-sics.mc\-mas\-ter\-.ca/Glo\-bu\-lar.html.
Previous information on these clusters can be summarized as follows.

{\bf NGC~6528} is the best studied cluster of our sample, due to its
moderate reddening E($B-V$)=0.52. Its overall metallicity is expected
to be [M/H]$\approx-0.2\div 0$.  Indeed, the strong similarity between
its optical CMD and that of NGC~6553 (Ortolani et al. 1995) suggests that the
two clusters have essentially the same metallicity. For NGC~6553
Barbuy et al. (1999) found [Fe/H]$\approx -0.55$ from the analysis of
two red giant stars, while Cohen et al. (1999) found [Fe/H] $\approx
-0.17$ from a sample of horizontal branch (HB) stars. Despite the
discrepancy in the measured [Fe/H] abundances, the two groups found
global metalliticy values that are compatible within the errors:
[M/H]$\approx -0.25$ according to Barbuy et al. (1999), and
[M/H]$-0.1\div +0.16$ (depending upon the adopted [O/Fe]) according to
Cohen et al. (1999).
Near-IR photometry for this cluster has been recently published by
Davidge (2000), but it does not reach the main sequence turnoff.

{\bf Terzan 4} appeared to have a blue horizontal branch (HB)
morphology, detected in a $V,I$ CMD (Ortolani et al. 1997a), obtained
under exceptional seeing conditions ($0\secondip34$). The results
suggested a metal-poor bulge cluster.

{\bf Terzan 5} is among the most metal-rich globular clusters in the
Galaxy, as indicated by its $V$,$I$ CMD (Ortolani et al. 1996a), and
integrated spectra (Bica et al. 1998). The CMD also shows strong
differential reddening effects.

{\bf Liller 1} is also quite metal rich, Frogel et al. (1995) derived
[Fe/H] = +0.25\-$\pm$0.3 from $J$, $H$ and $K$ CMDs. Further evidence
of a high metallicity was present in $I$ and Gunn $z$ CMDs of the Red
Giant Branch (RGB) (Ortolani et al. 1996b). A nice near-IR CMD of
Liller~1, reaching the turnoff magnitude, has been recently obtained by
Davidge (2000), but the strong contamination by bulge stars does not 
allow a reliable measure of the cluster turnoff magnitude.

{\bf UKS 1} is the most reddened of the program clusters, only the RGB
was so far detected by means of $I, z, J, K$ CMDs (Minniti et
al. 1995, Ortolani et al. 1997b).  A near-IR integrated spectrum
indicates a high metallicity (Bica et al. 1998), and infrared
echelle data (Rich et al. 2001) give [Fe/H]=$-$0.3 with
$\alpha/Fe=+0.3$.

The main purpose of the present paper is to use the minimized reddening
effects in the near-IR, combined with the  HST/NICMOS image quality to
produce highly improved CMDs, helping to determine the cluster ages.

In Section 2 the observations and reductions are described.  In Section 3
the CMDs of the clusters are presented. In Section 4 their ages and
other properties are discussed. Concluding remarks are given in
Section 5.

\begin{figure}
\centerline{\psfig{figure=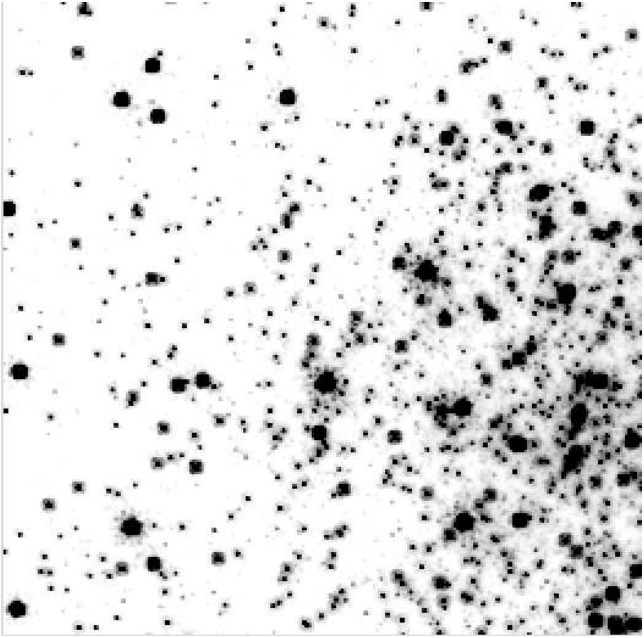,width=8.5cm}}
\caption{Terzan 5: F110W NIC2 offset image of the cluster.
Dimensions are 19.2"$\times$19.2".}
\label{field}
\end{figure}

\begin{table}
\begin{center}
\caption[1]{Log of observations}
\begin{tabular}{cccccc}
\hline
\noalign{\smallskip}
Cluster & Camera & Field & Filter & Date & Exp.(s)  \\
\noalign{\smallskip}
\hline
\noalign{\smallskip}
Terzan 4 & NIC2 & CEN & F110W & 27-04-98 & 256 \\
   "     &  "   & CEN & F160W & 27-04-98 & 320 \\
   "     &  "   & OFF & F160W & 27-04-98 & 352 \\
   "     &  "   & OFF & F110W & 27-04-98 & 256 \\
   "     & NIC1 & CEN & F110W & 27-04-98 & 256 \\
   "     &  "   & CEN & F160W & 27-04-98 & 320 \\
   "     &  "   & OFF & F110W & 27-04-98 & 256 \\
   "     &  "   & OFF & F160W & 27-04-98 & 352 \\
Terzan 5 & NIC2 & CEN & F110W & 23-04-98 & 256 \\
   "     &  "   & CEN & F160W & 23-04-98 & 320 \\
   "     &  "   & OFF & F110W & 23-04-98 & 256 \\
   "     &  "   & OFF & F160W & 23-04-98 & 352 \\
   "     & NIC1 & CEN & F110W & 23-04-98 & 256 \\
   "     &  "   & CEN & F160W & 23-04-98 & 320 \\
   "     &  "   & OFF & F110W & 23-04-98 & 256 \\
   "     &  "   & OFF & F160W & 23-04-98 & 352 \\
NGC 6528 & NIC2 & CEN & F110W & 17-04-98 & 256 \\
   "     &  "   & CEN & F160W & 17-04-98 & 320 \\
   "     &  "   & OFF & F110W & 17-04-98 & 256 \\
   "     &  "   & OFF & F160W & 17-04-98 & 352 \\
   "     & NIC1 & CEN & F110W & 17-04-98 & 256 \\
   "     &  "   & CEN & F160W & 17-04-98 & 320 \\
   "     &  "   & OFF & F110W & 17-04-98 & 256 \\
   "     &  "   & OFF & F160W & 17-04-98 & 352 \\
UKS 1    & NIC2 & CEN & F110W & 06-03-98 & 256 \\
   "     &  "   & CEN & F160W & 06-03-98 & 320 \\
   "     &  "   & OFF & F110W & 06-03-98 & 256 \\
   "     &  "   & OFF & F160W & 06-03-98 & 352 \\
   "     & NIC1 & CEN & F110W & 06-03-98 & 256 \\
   "     &  "   & CEN & F160W & 06-03-98 & 320 \\
   "     &  "   & OFF & F110W & 06-03-98 & 256 \\
   "     &  "   & OFF & F160W & 06-03-98 & 352 \\
Liller 1 & NIC2 & CEN & F110W & 08-08-98 & 256 \\
   "     &  "   & CEN & F160W & 08-08-98 & 320 \\
   "     &  "   & OFF & F110W & 08-08-98 & 256 \\
   "     &  "   & OFF & F160W & 08-08-98 & 352 \\
   "     & NIC1 & CEN & F110W & 08-08-98 & 256 \\
   "     &  "   & CEN & F160W & 08-08-98 & 320 \\
   "     &  "   & OFF & F110W & 08-08-98 & 256 \\
   "     &  "   & OFF & F160W & 08-08-98 & 352 \\
\noalign{\smallskip} \hline \end{tabular}
\end{center} 
\end{table}

\section{Observations and reductions}

The clusters were observed with NICMOS on board HST, through the F110W
and F160W filters.  Images were taken with NIC1 and NIC2 cameras. The
centers of the clusters were in NIC2, used as a primary camera, while
parallel observations with NIC1, which points $\sim 30$ arcsec away
from NIC2, provided higher resolution on the clusters more external
parts.  A second pointing of both cameras was taken with NIC2 located
at $\sim 20$ arcsec from the center of each cluster, in order to
sample less crowded regions.  All exposures were obtained using the
MULTIACCUM readout mode and the STEP64 time sequence through four
dithering positions.  Exposure times are reported in the log of
observations (Table 2).

The pixel size of NIC2 is $0\secondip075$, giving a field of view of
$19\secondip2\times19\secondip2$ for each frame, while NIC1 samples
$0\secondip043$ per pixel, for a total field of $11''\times11''$.  In
Fig. 1 we show, as an example, a F110W NIC2 image of Terzan 5. Note
the high angular resolution obtained.

The images were preprocessed by the standard NICMOS pipelines CALNICA
and CALNICB. These routines produce a single background-subtracted
mosaic image expressed in counts/sec/pixel. DAOPHOT II and ALLSTAR
(Stetson 1987) were used to determine the instrumental magnitudes. The
typical FWHM of the stars was $0\secondip11$. We determined the aperture
corrections from isolated stars and applied them to derive the stellar
magnitudes within a $0\secondip5$ aperture and normalized to nominal infinite
aperture.  The magnitudes were then converted to count rates and
multiplied by the inverse sensitivity given as the keyword PHOTFLAM 
in the header of the images. Finally, the zero point PHOTZPT was
applied to transform instrumental magnitudes into the standard HST 
m$_{110}$ and m$_{160}$ magnitudes used hereafter.

The presently available empirical transformations between NICMOS/HST
m$_{110}$ and m$_{160}$ and standard ground-based $J$ and $H$
(Stephens et al.\ 2000) are not suitable for the wide range of colours
covered by our observations. Instead, we prefer to use the theoretical
transformations derived by Origlia \& Leitherer (2000) to convert
theoretical isochrones into the HST/NICMOS observational plane.

\section{Colour Magnitude Diagrams}

\begin{figure}
\centerline{\psfig{figure=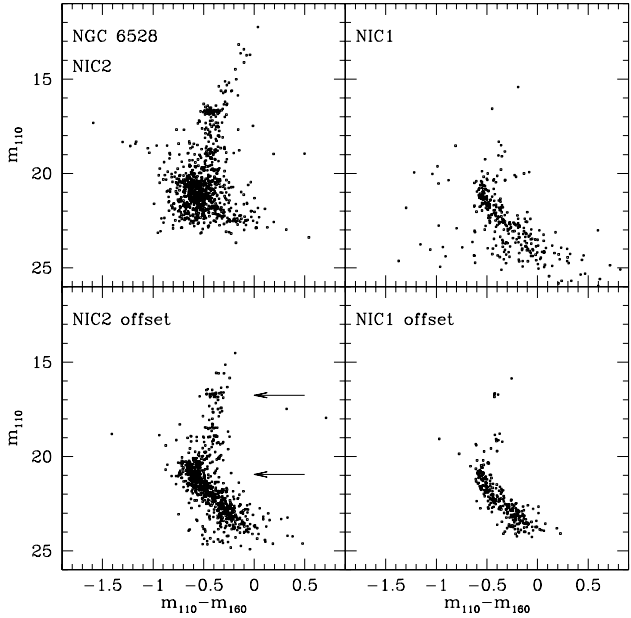,width=9.5cm}}
\caption{CMD of the four fields observed in NGC~6528. The upper left
panel refers to the cluster center region. The arrows show the
estimated positions of HB and turnoff.}
\label{n65}
\end{figure}

\begin{figure}
\centerline{\psfig{figure=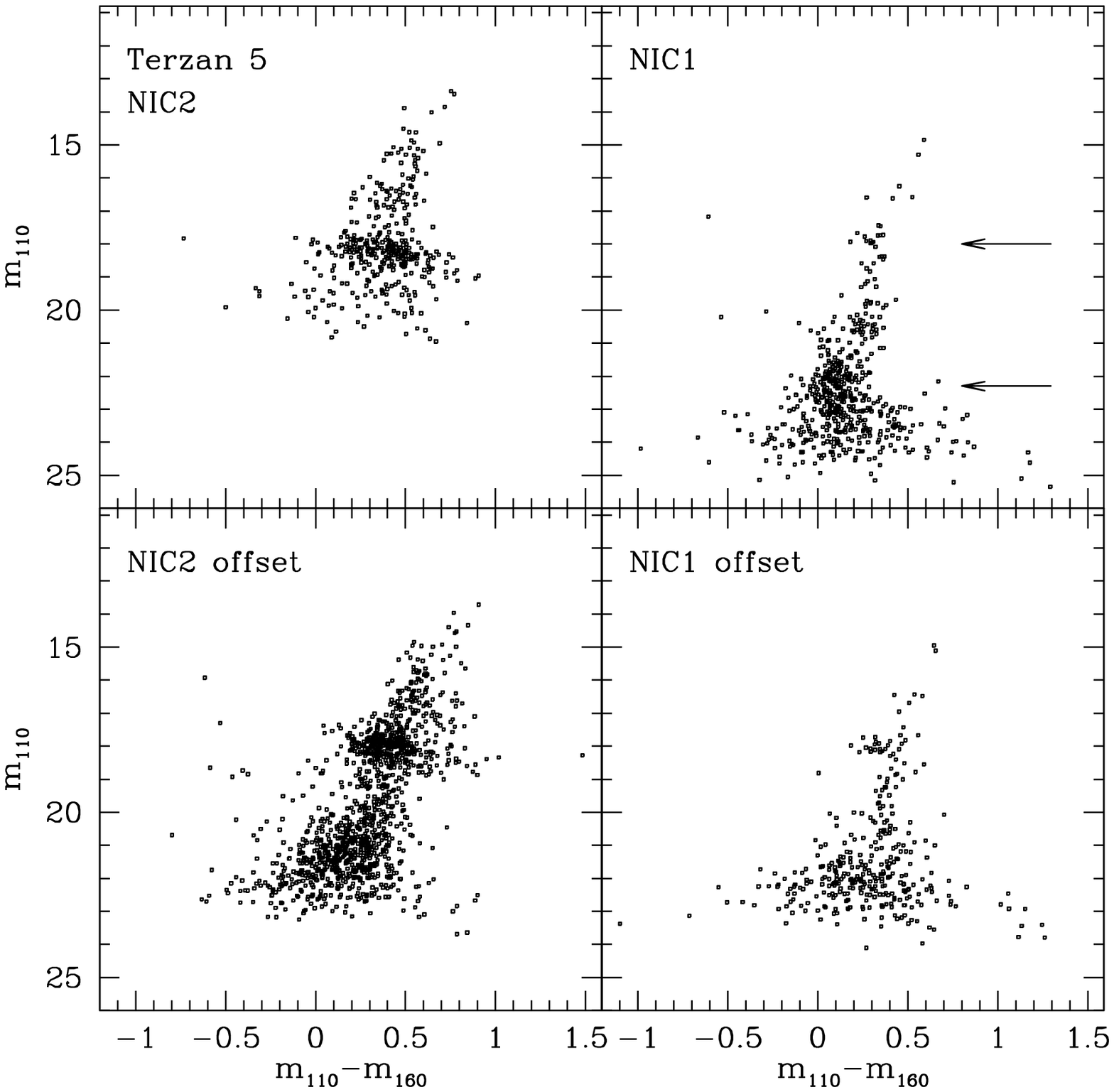,width=9.5cm}}
\caption{CMD of the four fields observed in Terzan~5. The upper left
panel refers to the cluster center region. The arrows
show the estimated positions of HB and turnoff.}
\label{tz5}
\end{figure}

\begin{figure}
\centerline{\psfig{figure=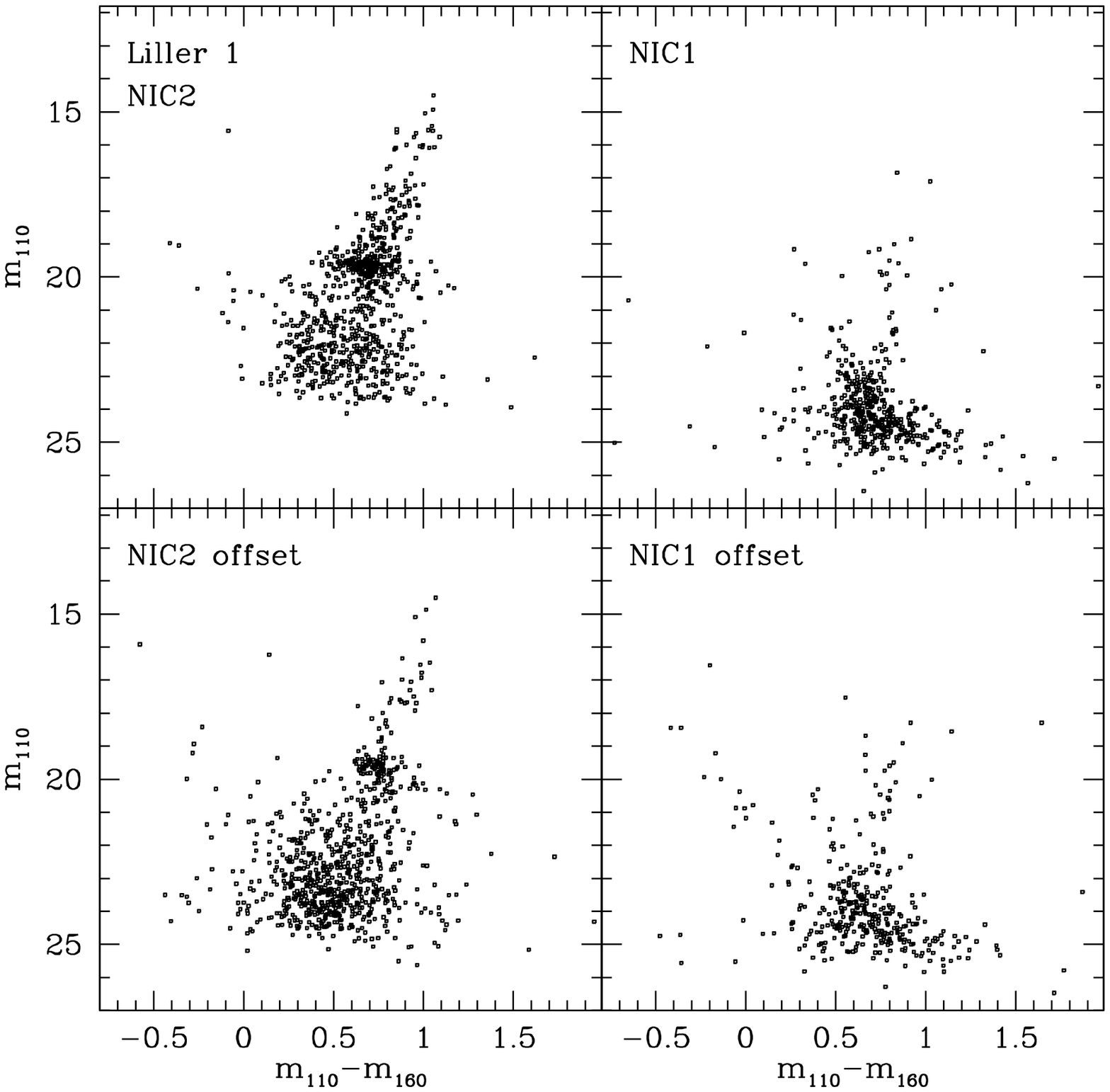,width=9.5cm}}
\caption{CMD of the four fields observed in Liller~1. The upper left
panel refers to the cluster center region.
}
\label{li1}
\end{figure}

\begin{figure}
\centerline{\psfig{figure=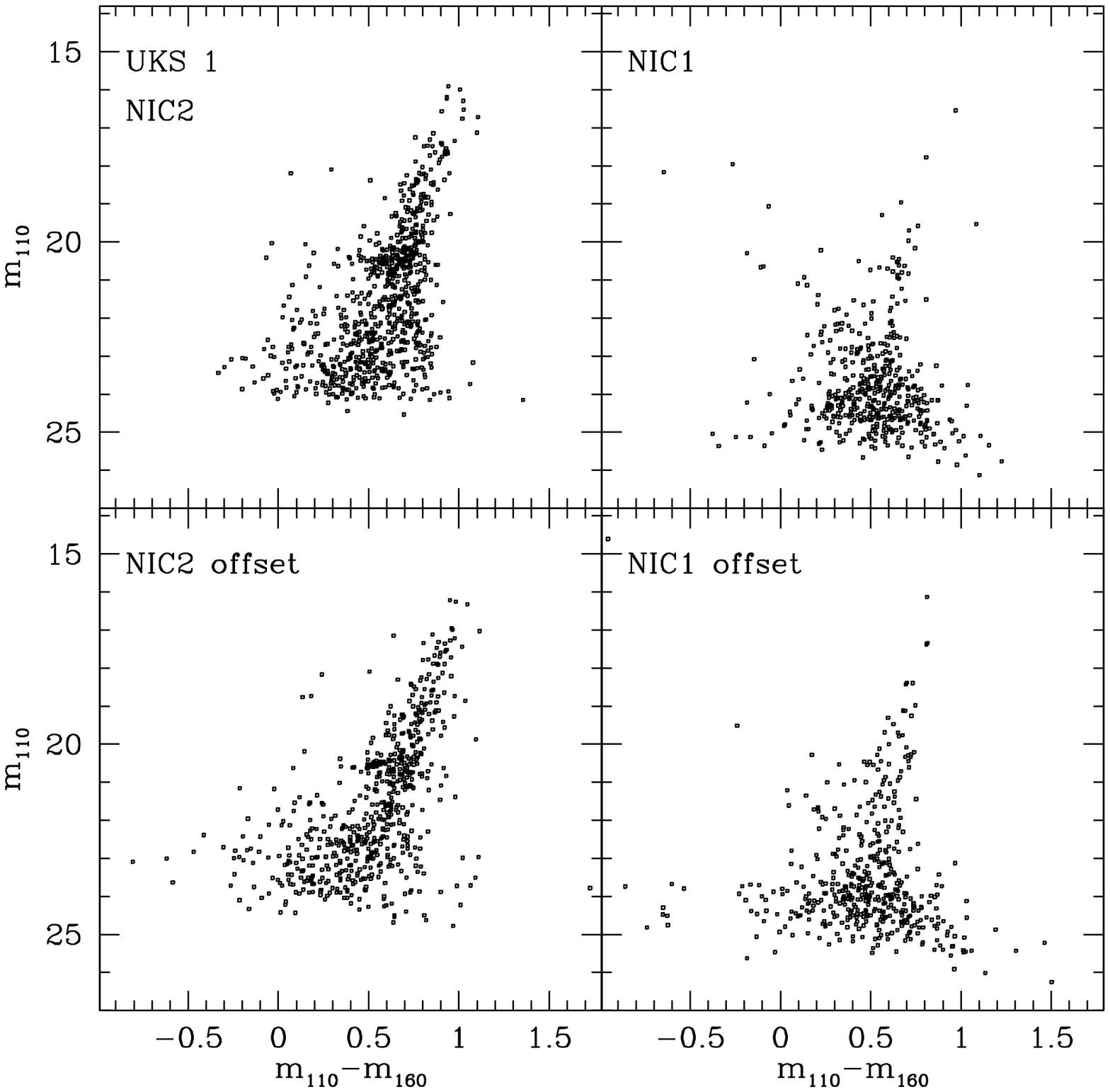,width=9.5cm}}
\caption{CMD of the four fields observed in UKS~1. The upper left
panel refers to the cluster center region.
}
\label{uks1}
\end{figure}

\begin{figure}
\centerline{\psfig{figure=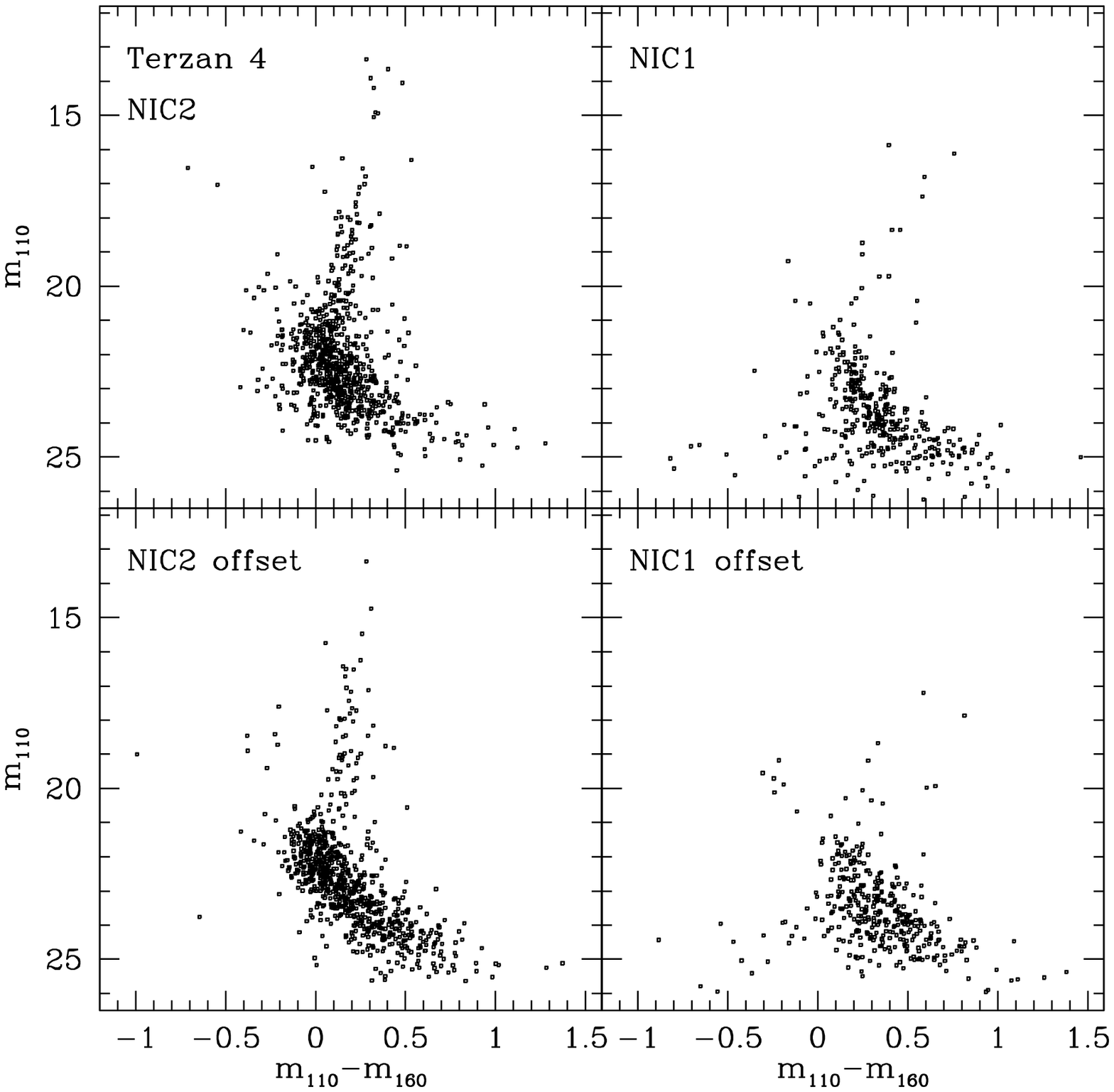,width=9.5cm}}
\caption{CMD of the four fields observed in Terzan~4. The upper left
panel refers to the cluster center region.}
\label{tz4}
\end{figure}

Figures 2 to 6 show the m$_{110}$ vs. m$_{110}$-m$_{160}$ CMDs of
NGC~6528, Terzan~5, Liller~1, UKS~1 and Terzan~4.  Each figure
contains 4 CMDs of a given cluster, corresponding to the four
pointings described above and indicated in the figure labels. NIC2 has
a wider field so that the corresponding panels show more populated
cluster sequences, in particular the HB. On the other hand, the NIC1
camera allowed us to obtain more accurate photometry and deeper
diagrams due to its smaller pixel size, giving a better PSF sampling,
hence reaching fainter magnitudes, especially in the most crowded
regions.

The CMDs of NGC~6528 (Fig. 2) reach more than 3 magnitudes below the 
turnoff in the two NIC1 and the NIC2-offset pointings.
 
The deepest CMD for Terzan~5, reaching about 1 mag below the turnoff,
is given in the upper right panel of Fig.~3. The brighter cluster
sequences are well populated in the NIC2 CMDs. The reduced effects of
differential reddening are noticeable with respect to the optical CMDs
(Ortolani et al. 1996a).

The red HB morphology of Liller~1 (Fig. 4) is clearly seen for the
first time.  Moreover the NIC1 CMDs have shown the cluster turnoff,
which is an important result for such a low Galactic latitude and very
reddened bulge cluster.  The red HB confirms the high metallicity,
as also derived by Frogel et al. (1995) from the RGB slope in the 
infrared CMD.

The HB morphology of UKS~1 (Fig. 5) is seen for the first time. The red
HB confirms that the cluster is metal rich (Bica et al. 1998). Since
the reddening values of UKS~1 and Liller~1 are quite similar (see
references in Sect. 1), the fainter HB of UKS~1 implies that it is more
distant. As a result the UKS~1 CMDs barely reach the SGB region.

Terzan~4 (Fig. 6) CMDs are very deep attaining more than 3 magnitudes
below the turnoff in both cameras. Note the absence of a red HB,
suggesting presence of a sparsely populated blue HB, which, in these
bands, would be located at m$_{110}$-m$_{160}\sim -0.3$ (see the two
lower panels of Fig. 6).  The metal poor nature of this cluster
(Ortolani et al. 1997a) is therefore confirmed.

\section{Cluster Ages} 

The magnitude difference between the HB and the turnoff
$\Delta$V$^{\rm HB}_{\rm TO}$ is an effective age indicator (e.g.,
Iben 1974; Sandage 1986), and the method has been widely used,
especially to derive relative ages (e.g. Ortolani et al. 1995, 1999;
Chaboyer et al. 1996; Rosenberg et al. 1999). In particular, the
method was applied to constrain in a fully empirical way any age
difference between the bulge globulars NGC~6528 and NGC~6553 and the
halo globular clusters (Ortolani et al. 1995).

While most of these studies used optical photometry, we now adopt the
same approach using the NICMOS near-IR data.  From the CMDs presented
in the previous section accurate values of $\Delta m_{110}
=m_{110}^{\rm HB}- m_{110}^{\rm TO}$ can be obtained only for
NGC 6528 and Terzan~5.  In Figs. 2 and 3 the HB and turnoff locations
are indicated by arrows.  The resulting $\Delta m_{110}$ are reported
in Table~3 together with the estimated uncertainty.  Unfortunately,
the CMDs for Liller~1 and UKS~1 are not accurate enough for a similar
measure: the turnoff is not very well defined, and it might appear
brighter than it is due to crowding (blends). Moreover, for these
clusters the turnoff and the HB were not measurable from a single
camera, and possible residual offsets between NIC1 and NIC2 may
introduce a systematic error. In the case of Terzan 4 the turnoff is
fairly well defined, but the blue morphology together with the sparse
population of the HB makes it impossible to derive the corresponding value
of $\Delta m_{110}$.  The value of $\Delta m_{110}$ for 47~Tuc was
also measured from a CMD obtained from NIC3 archive data (program
GO7318) taken through the same filters. The result is also reported in
Table 3 and will be used later to constrain the age difference, if any,
between the bulge clusters and this inner halo cluster.  The overall
metallicity of 47~Tuc was adopted to be [M/H]=$-0.5$, given its iron
abundance [Fe/H]$=-0.7$ and $\alpha$-element enhancement
[$\alpha$/Fe]$\simeq 0.3$ (Gratton, Quarta \& Ortolani 1986).

\subsection{Isochrones in the NICMOS m$_{110}$-m$_{160}$ System}

To further constrain the ages of the program clusters new
iso\-chro\-nes in the NICMOS photometric system were constructed.
All the evolutionary tracks used for producing the set of isochrones
adopted in the present analysis have been obtained by means of the
FRANEC evolutionary code (Cassisi \& Salaris 1997, Castellani et
al. 1999).  The input physics includes the OPAL radiative opacities
(Iglesias \& Rogers 1996) for temperatures higher than 10.000 K, while
for lower temperatures the molecular opacity tables provided by
Alexander \& Ferguson (1994) have been adopted. This choice allows us
to have a smooth match between the two different opacity sets.  Both
high and low-temperature opacities have been computed by assuming a
solar-scaled heavy elements distribution (Greves\-se 1991).  The
Straniero (1988) equation of state was adopted,
while superadiabatic convection was treated in the mixing length
approximation with calibration provided by Salaris \& Cassisi
(1996). More on the input physics can be found in Cassisi \& Salaris
(1997). The evolutionary computations do not consider atomic
diffusion. However, it is well known (Castellani et al. 1997, Cassisi
et al. 1998) that accounting for such a mechanism leads to a 
reduction of the cluster ages by $\sim 0.7-1$ Gyr.

Stellar models and isochrones have been constructed for the following
initial chemical abundances: $(Z=0.006, Y=0.25)$ and $(Z=0.02, Y=0.289)$,
which assume a helium enrichment ratio $dY/dZ=2.8$.  For
each adopted chemical composition, Zero Age Horizontal Branch (ZAHB)
models corresponding to a RGB progenitor with initial mass equal to
$1M_\odot$ have been computed. In fact, the ZAHB structures have been
constructed for each fixed metallicity by using the helium core mass
and the envelope chemical abundance profile suitable for a RGB
progenitor with this mass.  All the isochrones have been transferred
from the theoretical plane to the observational one in the HST/NICMOS
photometric system (F110W, F160W and F222W filters) using recent
prescriptions (Origlia \& Leitherer 2000).  The resulting
$\Delta m_{110}$ values for ages in the range 7 $\leq t\leq$ 17 (Gyr) 
are listed in Table~3.

\subsection{Results}

\begin{table}
\begin{center}
\caption[1]{Comparison of $\Delta m_{110}$ measured 
on the CMDs and in theoretical isochrones. A ratio $dY/dZ=2.8$
has been assumed in these models, see discussion in the text.}
\begin{tabular}{lrcrrc}
\hline
\noalign{\smallskip}
Object &  [M/H]  &  Age (Gyr)& $m_{110}^{\rm HB}$ & $m_{110}^{\rm TO}$ &
$\Delta m_{110}$ \\
\noalign{\smallskip}
\hline
\noalign{\smallskip}
47~Tuc    &  $-0.50$ & $15\pm3$ & 16.50 & 20.70 & 4.2$\pm$0.20  \\
Terzan~5  &  $ 0.00$ & $14\pm3$ & 18.00 & 22.30 & 4.3$\pm$0.20  \\
NGC~6528  &  $-0.20$ & $13\pm3$ & 16.75 & 20.95 & 4.2$\pm$0.15  \\
Terzan~4  &  $-2.00$ &          &       & 21.90 &   		\\
Liller~1  &  $+0.20$ &          & 19.60 &       &   		\\
UKS~1     &  $-0.50$ &          & 20.50 &       &   		\\
\noalign{\smallskip}
model     &   0.00   &  10      &  	&  	&	   4.13 \\
model     &   0.00   &  12      &  	&  	&	   4.22 \\
model     &   0.00   &  14      &  	&  	&	   4.30 \\
model     &   0.00   &  16      &  	&  	&	   4.37 \\
\noalign{\smallskip}
model     & $-0.50$  &  10      &       &	&	   3.86 \\
model     & $-0.50$  &  12      &       &	&	   4.06 \\
model     & $-0.50$  &  14      &       &	&	   4.14 \\
model	  & $-0.50$  &  15	& 	&       &	   4.20 \\
model     & $-0.50$  &  16      &       &	&	   4.25 \\
model     & $-0.50$  &  17      &       &	&	   4.30 \\
\noalign{\smallskip} \hline \end{tabular}
\end{center} 
\end{table}

We first proceed differentially, comparing the $\Delta m_{110}$ values
of the program clusters to that of the reference cluster 47~Tuc. As reported 
in Table~3, the values of $\Delta m_{110}$ for Terzan~5 and NGC~6528 are
identical to the value for 47~Tuc, well within the $\pm 0.2$ mag uncertainty.
There are, however, metallicity differences by $\sim 0.2$ and $\sim 0.3$ dex
among the three clusters that must be taken into account. From the theoretical
values also listed in Table~3 we see that a 0.25 dex difference in metallicity
results in a variation by $\sim 0.08$ mag in $\Delta m_{110}$, at fixed age.

As a rule of thumb, to a good approximation cluster relative age
errors (variations) are identical to errors (variations) in the
turnoff magnitude, i.e., $\delta t/t \approx \delta M_{TO}$ (e.g.,
Renzini 1991). Since the HB luminosity is virtually age
independent, this implies that cluster age errors (variations) are
identical to errors (variations) in $\Delta m_{110}$.  Having
determined the distance to 47~Tuc using the white dwarf method,
Zoccali et al. (2001b) derive an age of 13$\pm 2.5$ Gyr and 14 $\pm
2.5$ Gyr for this cluster, using respectively models with and without
diffusion. Given the values of $\Delta m_{110}$ reported in Table~3,
and that any effect of metallicity differences among the clusters is
obliterated by the error on the $\Delta m_{110}$ values, we conclude
that the three clusters are coeval, within a $\pm20\%$ uncertainty, or
13 (14) $\pm 2.6$ Gyr, the value in parenthesis referring to models
neglecting diffusion.

As a second approach, we proceed using the full theoretical 
age-metallicity-$\Delta m_{110}$ relation as reported in Table~3. By 
interpolating in both $\Delta m_{110}$ and [M/H] one then derives an age of
15, 15.5, and 13 Gyr, respectively for 47~Tuc, Terzan~5, and NGC~6528.
Being all affected by a $\sim 20\%$ uncertainty, we do not consider as
significant these age differences, and once more conclude that no evidence
exists for an age difference between these halo and bulge clusters.

These results depend somewhat on the adopted helium-enrichment ratio
($dY/dZ\approx2.8$).  However, the exact value of this parameter is
still a matter of debate.  Recent works by Sandquist (2000) and
Zoccali et al. (2000) on helium indicators in galactic globular
clusters suggest that $dY/dZ<2.5$.  Given that, for each fixed cluster
age, the luminosity of both the turnoff and the ZAHB strongly depend
on the He abundance, we have estimated how much the present results on
the cluster ages are affected by our choice on $dY/dZ$.  By performing
several numerical experiments, we have verified that, at solar
metallicity and ages of the order of 10~Gyr, the difference in
$m_{110}$ magnitude between the turnoff and the reddest point along
the theoretical ZAHB depends on the initial He content as: $\Delta
m_{110}/dY\approx2.9$.  This means that the $\Delta m_{110}$ values
shown in Table~3, for solar metallicity, have to be decreased by about
0.11 mag at each fixed age, if an initial He content Y=0.25 is 
assumed, i.e., the same adopted for the isochrones for the models at
metallicity $[M/H]=-0.5$. This would increase the age of NGC~6528 and
Terzan~5 to about $14\pm3$ and $16\pm3$ Gyr, respectively.

It is worth recalling that the {\it absolute} ages determined here 
may be overestimated by $\sim 1$ Gyr, due to the fact that
the HB level has been measured in the observed CMD as the mean locus
of the HB stars, which is about $\approx 0.1$ mag brighter than the
ZAHB level given by the theoretical models. We did not try to simulate
HB evolution in our models, nor to estimate the lower envelope of the
empirical HB distribution because in any case the magnitude spread of
the observed HB, mainly due to differential reddening, does not allow
precision in the ages better than $\sim 3$ Gyr.

Despite this, the data constrain the age of these clusters
to be old (like 47 Tuc), compatible with halo globular cluster ages and
confirming the previous HST optical results on the metal-rich bulge
clusters NGC~6528 and NGC~6553 (Ortolani et al. 1995).
   
\section{Concluding remarks}

The magnitude difference between the horizontal branch and the turnoff
$\Delta m_{110}$, combined with newly transformed stellar evolution
models, gives relative and absolute ages for the clusters 47 Tuc,
NGC~6528 and Terzan~5.  Within typical $\sim 20\%$ uncertainty, the
bulge clusters NGC 6528 and Terzan 5 appear to be coeval with 47 Tuc
and halo globular clusters. In the case of NGC 6528 the results by
Ortolani et al. (1995) based on HST V and I photometry are confirmed,
suggesting that the bulge is as old as NGC~6528 (and 47~Tuc).

The NICMOS CMDs presented here for the clusters Liller 1, UKS 1 and
Terzan 4 are the deepest so far obtained for these clusters that in
the optical are affected by 7-9 magnitudes of extinction. These
unprecedented observations reveal the red horizontal branch morphology
of Liller 1 and UKS 1, and reach the turnoff of these clusters, albeit
with insufficient accuracy to allow a meaningful determination of
$\Delta m_{110}$. Both the red horizontal branch morphology and the
slope of the RGB suggest that UKS 1 is a metal-rich cluster.  In the
less reddened cluster Terzan 4 the turnoff region is well delineated,
but its sparse population and blue horizontal branch prevent us from
estimating $\Delta m_{110}$. However we confirm that Terzan~4 has the 
blue horizontal branch  characteristic of a very metal poor cluster.

\begin{acknowledgements}

We acknowledge partial financial support from the Brazilian agencies
CNPq and Fapesp. SO acknowledges Italian Ministero dell'Universit\`a e
della Ricerca Scientifica e Tecnologica (MURST) under the program on
'Stellar Dynamics and Stellar Evolution in Globular Clusters: a
Challenge for New Astronomical Instruments'.  SC thanks for the
financial support by MURST - Cofin2000- under the scientific project
"Stellar Observables of Cosmological Relevance" (Italy). Support for
R.M.R. was provided by NASA through grant number GO-7832 from the
Space Telescope Science Institute, which is operated by AURA, Inc.,
under NASA contract NAS5-26555.

\end{acknowledgements}

%

\end{document}